\documentclass[pre,twocolumn]{revtex4}
\usepackage{amsmath}
\usepackage{amssymb} 
\usepackage{amsbsy}
\usepackage{graphics}
\usepackage{color}
\usepackage{rotating}

\begin{document}

\title{An extended analysis of the viscosity kernel for monatomic and diatomic fluids}

\author{R. M. Puscasu}
\email{rpuscasu@swin.edu.au}
\affiliation{Centre for Molecular Simulation, Swinburne University of
Technology, PO Box 218, Hawthorn, Victoria 3122, Australia }

\author{B. D. Todd}
\email{btodd@swin.edu.au}
\affiliation{Centre for Molecular Simulation, Swinburne University of
Technology, PO Box 218, Hawthorn, Victoria 3122, Australia  }

\author{P. J. Daivis}
\email{peter.daivis@rmit.edu.au}
\affiliation{Applied Physics, School of Applied Sciences, RMIT University, 
G.P.O. Box 2476, Melbourne, Victoria 3001, Australia}

\author{J. S. Hansen}
\email{jehansen@swin.edu.au}
\affiliation{Centre for Molecular Simulation, Swinburne University of
Technology, PO Box 218, Hawthorn, Victoria 3122, Australia}

\date{\today} 

\begin{abstract}
\vspace{5pt}
We present an extended analysis of the wave-vector dependent shear viscosity of monatomic and diatomic (liquid chlorine) fluids over a wide range of wave-vectors and for a variety of state points. The analysis is based on equilibrium molecular dynamics simulations, which involves the evaluation of transverse momentum density and shear stress autocorrelation functions. For liquid chlorine we present the results in both atomic and molecular formalisms. We find that the viscosity kernel of chlorine is statistically indistinguishable with respect to atomic and molecular formalisms. The results further suggest that the real space viscosity kernels of monatomic and diatomic fluids depends sensitively on the density, the potential energy function and the choice of fitting function in reciprocal space. It is also shown that the reciprocal space shear viscosity data can be fitted to two different simple functional forms over the entire density, temperature and wave-vector range: a function composed of \textit{n}-Gaussian terms and a Lorentzian type function. Overall, the real space viscosity kernel has a width of 3 to 6 atomic diameters which means that the generalized hydrodynamic constitutive relation is required for fluids with strain rates that vary nonlinearly over distances of the order of atomic dimensions.
\vspace{15pt}
\end{abstract}


\maketitle

\section{Introduction \label{sect:intro}}

Fluid dynamics at atomic and molecular scales still present a challenge for theoreticians as well as for experimentalists. Molecular dynamics (MD) is a computational tool that has contributed significantly to the fundamental understanding of these systems by providing information about processes not directly approachable by experimental studies. A central problem in the study of fluids at such small length and time scales is the computation of meaningful transport properties. Many equilibrium molecular dynamics (EMD) as well as nonequilibrium molecular dynamics (NEMD) simulations of nanofluids have been performed since the early 1980s \cite{bitsanis_1988,bitsanis_1990,oconnell_1995,koplik_1995,nie_2003}. In most of these simulations the stress was treated as being dependent of the local strain rate rather than the entire strain rate distribution in the system. Todd \textit{et al.} have recently shown that in all but the simplest flows (e.g. planar Couette and Poiseuille flows) and for velocity fields with high gradients in the strain rate over the width of the real space viscosity kernel, non-locality can play a significant role \cite{todd_2008,todd_2008_a}. In the case of a homogeneous fluid, a local viscosity defined by Newton's viscosity law as
\begin{equation}
P_{xy}(\mathbf{r},t)= -\eta_{0}\int\limits_{0}^{t} \int\limits_{-\infty}^{\infty} \delta(\mathbf{r}-\mathbf{r}',t-t')\dot{\gamma}(\mathbf{r}',t') d\mathbf{r}'dt' \label{eqn:homo_localceq}
\end{equation}
even exhibits singularities at points where the strain rate is zero \cite{travis_1997,travis_2000,zhang_2004,zhang_2005}. In Eq.~(\ref{eqn:homo_localceq}) $P_{xy}(\mathbf{r},t)$ represents the $(x,y)$ off-diagonal component of the pressure tensor, $\dot {\gamma}(\mathbf{r},t)$ is the shear strain rate at position $\mathbf{r}$ and time $t$, and $\eta_{0}$ is the local shear viscosity. In general, a nonlocal constitutive equation that allows for spatial and temporal non-locality can be expressed as \cite{alley_1983,evans_1990}  
\begin{equation}
P_{xy}(\mathbf{r},t)= -\int\limits_{0}^{t} \int\limits_{-\infty}^{\infty}\eta(\mathbf{r}-\mathbf{r}',t-t')
\dot{\gamma}(\mathbf{r}',t') d\mathbf{r}'dt' , \label{eqn:homo_nonlocalceq}
\end{equation}
for a homogeneous fluid. In the situation where the strain rate is constant in time and only varies with respect to the spatial coordinate $y$, Eq.~(\ref{eqn:homo_nonlocalceq}) can be written as 
\begin{equation}
P_{xy}(y)= -\int\limits_{-\infty}^{\infty} \eta(y-y')\dot{\gamma}(y') dy' .
\label{eqn:srcttime_homo_nonlocalceq}
\end{equation}
In reciprocal space Eq.~(\ref{eqn:srcttime_homo_nonlocalceq}), can be expressed as
\begin{equation}
\tilde{P}_{xy}(k_{y})= - \tilde{\eta}(k_{y}) \tilde{\dot{\gamma}}(k_{y}) ,
\label{eqn:kspace_homo_nonlocalceq}
\end{equation}
where $k_{y}$ is the $y$ component of the wave-vector as defined later in Section \ref{sect:III}. Such a constitutive equation is expected to be necessary for the description of flows in highly confined systems, due to the large change in the strain rate with position in the vicinity of the wall \cite{travis_1997}.

The best available theoretical predictions of the wave-vector dependent viscosity are based on mode-coupling theory and generalized Enskog theory \cite{leutheusser_1982_1,leutheusser_1982_2,yip_1982}. However, these theories do not quantitatively agree with data obtained via computer simulations \cite{alley_1983}. The theoretical predictions focus on the transverse momentum density autocorrelation function, which is found by an iterative numerical solution of a system of nonlinear equations. Consequently, the theories do not result in analytical expressions for the correlation functions or the wave-vector dependent transport coefficients, which are the focus of the present study. More recently, a modified collective mode approach has been successfully applied by Omelyan \textit{et al.} \cite{omelyan_2005} to the TIP4 model of water. In contrast to other semi-phenomenological approaches used for instance in TIP4P and SPC/E models of water by Bertolini \textit{et al.} \cite {bertolini_1995} and Palmer \cite{palmer_1994}, Omelyan \textit{et al.} reproduced the reciprocal space kernel using a relatively small number of modes. 

In this paper, we extend the work done by Hansen \textit{et al.} \cite{hansen_2007_2} and focus on computing the spatially non-local viscosity kernel for monatomic and diatomic fluids over a wider range of wave-vectors, state points and potential energy functions. We are specifically interested in identifying functional forms that fit the reciprocal space kernel data. On the basis of these results, we are be able to assess the length scale (i.e. the width of the real space kernel) over which the governing generalized constitutive relation Eq.~(\ref{eqn:srcttime_homo_nonlocalceq}) must be used. We expect that non-local transport phenomena would be relevant in shock waves \cite{alley_1983,holian_1980,holian_1998,reed_2006,reed_2003}, shear banding \cite{dhont_1999}, flows of micellar solutions \cite{masselon_2008}, suspensions of rigid fibers \cite{schiek_1995} and jammed or glassy systems \cite{bocquet_2008}.

This paper is structured as follows: In Section \ref{sect:IIa} we give an overview of the general formulation and the expressions for the complex wave-vector and frequency dependent viscosity are given. In Section \ref{sect:III} we describe the simulation methodology and conditions. In Section \ref{sect:IV} we present our molecular dynamics simulation data and compare the results of our monatomic and diatomic viscosity kernels, particularly the shape of the kernels. Finally, we summarize and conclude our analysis in Section 5.

\section{Methodology \label{sect:method}}

We shall briefly introduce the main conceptual background used in this work, namely, the wave-vector dependent momentum density, stress and viscosity in the atomic and molecular formulations.

\subsection{Wave-vector dependent momentum density for atomic and molecular fluids \label{sect:IIa}}

For a single component atomic fluid the real space microscopic momentum density is given by \cite{evans_1990}
\begin{equation}
\mathbf{J}(\mathbf{r},t)= \rho_{a}(\mathbf{r},t)\mathbf{v}(\mathbf{r},t)= \sum_{i=1}^{N_{a}} m_{i} \mathbf{v}_{i}(t)\delta(\mathbf{r}-\mathbf{r}_{i}) \label{eqn:af_atomic_momentum_density}
\end{equation}
where $\rho_{a}(\mathbf{r},t)=\sum_{i=1}^{N_{a}} m_{i} \delta(\mathbf{r}-\mathbf{r}_{i})$ is the mass density, $m_i$, $\mathbf{r}_{i}$ and $\mathbf{v}_{i}$ are the mass, position and velocity of atom $i$. The summation runs over the number of atoms $N_a$ in the system. The Fourier transform of the momentum density is
\begin{equation}
\tilde{\mathbf{J}}(\mathbf{k},t)= \sum_{i=1}^{N_{a}} m_{i} \mathbf{v}_{i}(t) e^{i\mathbf{k}\cdot\mathbf{r}_{i}} \label{eqn:af_atomic_momentum_density_kspace}
\end{equation}
while the Fourier transform of the mass density is $\tilde{\rho}_{a}(\mathbf{k},t)=\sum_{i=1}^{N_{a}} m_{i} e^{i\mathbf{k}\cdot\mathbf{r}_{i}}$. We define the Fourier transform of a function $f(r)$ as $\mathcal{F}[f(r)]=\tilde{f}(k)=\int_{-\infty}^{\infty}e^{ikr}f(r)dr$.

The atomic representation of the momentum density for a molecular fluid can be written in real space as \cite{todd_2007}:  
\begin{equation}
\mathbf{J}^{A}(\mathbf{r},t) = \rho_{a}(\mathbf{r},t)\mathbf{v}(\mathbf{r},t)=\sum_{i=1}^{N_{m}}\sum_{\alpha=1}^{N_{s}} m_{i\alpha}\mathbf{v}_{i\alpha}(t)\delta(\mathbf{r}-\mathbf{r}_{i\alpha}) \label{eqn:atomic_momentum_density}
\end{equation}
where the mass density is defined as $\rho_{a}(\mathbf{r},t)=\sum_{i=1}^{N_{m}}\sum_{\alpha=1}^{N_{s}} m_{i\alpha} \delta(\mathbf{r}-\mathbf{r}_{i\alpha})$. The inner summation extends over the $N_{s}$ mass points in a molecule and the outer summation extends over the number of molecules $N_{m}$ in the system. In general, $N_{s}$ depends on the molecule index $i$ for a multicomponent system, but in our systems $N_{s}$ is the same for all molecules and the interaction sites are assumed to have identical mass, namely $m_{i\alpha}$. 

The Fourier transform of the momentum density is
\begin{equation}
\tilde{\mathbf{J}}^{A}(\mathbf{k},t)= \sum_{i=1}^{N_{m}}\sum_{\alpha=1}^{N_{s}} m_{i\alpha} \mathbf{v}_{i\alpha}(t) 
e^{i\mathbf{k}\cdot\mathbf{r}_{i\alpha}}
\label{eqn:atomic_momentum_density_kspace}
\end{equation}
where the transformed mass density is $\tilde{\rho}_{a}(\mathbf{k},t)= \sum_{i=1}^{N_{m}}\sum_{\alpha=1}^{N_{s}} m_{i\alpha} e^{i\mathbf{k}\cdot\mathbf{r}_{i\alpha}}$. For molecules composed of $N_{s}$ atoms we can define the mass of molecule $i$,  $M_{i}=\sum_{\alpha=1}^{N_{s}} m_{i\alpha}$, position of the 
molecular center of mass as $\mathbf{r}_{i}=\sum_{\alpha=1}^{N_{s}} m_{i\alpha} \mathbf{r}_{i\alpha} / M_{i}$, position of site $\alpha$ of molecule $i$ relative to the center of mass of molecule $i$ as $\mathbf{R}_{i\alpha}=\mathbf{r}_{i\alpha}-\mathbf{r}_{i}$, and center of mass momentum of the molecule as
$\mathbf{p}_{i}=\sum_{\alpha=1}^{N_{s}} \mathbf{p}_{i\alpha}$. This means that the atomic mass density can be written in \textbf{k}-space as $\tilde{\rho}_{a}(\mathbf{k},t)=\sum_{i=1}^{N_{m}}\sum_{\alpha=1}^{N_{s}} m_{i\alpha}e^{i\mathbf{k}\cdot(\mathbf{r}_{i}+\mathbf{R}_{i\alpha})}$. If we expand this relation further we can express the atomic mass density in terms of molecular mass density in which we define the mass density in the molecular representation as $\tilde{\rho}_{m}(\mathbf{k},t)=\sum_{i=1}^{N_{m}}M_{i}e^{i\mathbf{k}\cdot \mathbf{r}_{i}}$ 
in reciprocal space and as $\rho_{m}(\mathbf{r},t)=\sum_{i=1}^{N_{m}}M_{i}\delta(\mathbf{r}-\mathbf{r}_{i})$ in real space, respectively. In a similar way we can expand the atomic momentum density about the molecular center of mass: $\tilde{\mathbf{J}}^{A}(\mathbf{k},t)=\sum_{i=1}^{N_{m}}\sum_{\alpha=1}^{N_{s}} m_{i\alpha} \mathbf{v}_{i\alpha}(1+i\mathbf{k}\cdot\mathbf{R}_{i\alpha}+\dots)e^{i\mathbf{k}\cdot\mathbf{r}_{i\alpha}}$.

The Fourier transform of the momentum density in the molecular representation can then be defined as
\begin{equation}
\tilde{\mathbf{J}}^{M}(\mathbf{k},t)= \sum_{i=1}^{N_{m}}M_{i} \mathbf{v}_{i}(t) e^{i\mathbf{k}\cdot\mathbf{r}_{i}} \label{eqn:atomic_momentum_density_kspace_2}
\end{equation}
A complete procedure for expressing the mass and momentum densities in physical and reciprocal space for atomic and molecular fluids has been discussed in more detail by Todd and Daivis \cite{todd_2007}.

\subsection{Wave-vector dependent pressure tensor \label{sect:IIb}}

For a monatomic system the wave-vector dependent pressure tensor is defined as \cite{todd_2007}
\begin{equation}
\tilde{\mathbf{P}}(\mathbf{k},t)= \sum_{i=1}^{N} \frac{\mathbf{p}_{i} \mathbf{p}_{i}}{m_{i}}e^{i\mathbf{k} \cdot \mathbf{r}_{i}}-\frac{1}{2} \sum_{i=1}^{N}\sum_{j\ne i}^{N} \mathbf{r}_{ij} \mathbf{F}_{ij}g(\mathbf{k}) e^{i\mathbf{k} \cdot \mathbf{r}_{i}} \label{eqn:atomic_atomic_pressure_tensor}
\end{equation}
where $\mathbf{F}_{i}$ is the force on atom $i$ due to atom $j$ and $g(\mathbf{k})=(e^{i\mathbf{k} \cdot \mathbf{r}}-1)/i\mathbf{k} \cdot \mathbf{r}=\sum_{n=0}^{\infty}(i\mathbf{k}\cdot \mathbf{r})^{n}/{(n+1)!}$ is the Fourier transform of the Irving-Kirkwood $O_{ij}$ operator  \cite{irving_1950}.

For a molecular system the molecular pressure tensor is the pressure calculated using the intermolecular forces and the molecular center of mass momenta. The atomic pressure on the other hand includes all atomic momenta and all interatomic forces and constraint forces. Thus the wave-vector dependent pressure tensor for constrained diatomic fluid can be written in atomic representation as 
\begin{eqnarray}
\tilde{\mathbf{P}}^{A}(\mathbf{k},t)&=&\sum_{i=1}^{N_m}\sum_{\alpha=1}^{2} \frac{\mathbf{p}_{i\alpha} \mathbf{p}_{i\alpha}}{m_{i\alpha}}e^{i\mathbf{k} \cdot \mathbf{r}_{i\alpha}} \nonumber \\
&-&\frac{1}{2} \sum_{i=1}^{N_m} \sum_{\alpha=1}^{2} \sum_{j\neq i}^{N_m} \sum_{\beta=1}^{2} \mathbf{r}_{i\alpha j\beta} \mathbf{F}_{i\alpha j\beta} 
g(\mathbf{k}) e^{i\mathbf{k} \cdot \mathbf{r}_{i\alpha j\beta}} \nonumber \\
&+&\sum_{i=1}^{N_m} \sum_{\alpha=1}^{2} \mathbf{r}_{i\alpha} \mathbf{F}_{i\alpha}^{C} 
g(\mathbf{k}) e^{i\mathbf{k} \cdot \mathbf{r}_{i\alpha}}
\label{eqn:atomic_pressure_tensor}
\end{eqnarray}
where $\mathbf{F}_{i\alpha j \beta}$ is the LJ force acting on site $\alpha$ of molecule $i$ due to site $\beta$ of molecule $j$, and  $\mathbf{F}_{i\alpha}^{C}$ is the bond constraint force on site $\alpha$ of molecule $i$. $\mathbf{r}_{i\alpha j \beta} =\mathbf{r}_{j\beta}-\mathbf{r}_{i\alpha}$ is the minimum image separation of site $\alpha$ of molecule $i$ from site $\beta$ of molecule $j$. 

The pressure tensor for a diatomic molecule in the molecular representation is defined as
\begin{eqnarray}
\tilde{\mathbf{P}}^{M}(\mathbf{k},t)= \sum_{i=1}^{N_m} \frac{\mathbf{p}_{i} \mathbf{p}_{i}} {M_{i}}e^{i\mathbf{k} \cdot \mathbf{r}_{i}} -\frac{1}{2} \sum_{i=1}^{N_m} \sum_{j\ne i}^{N_m}  \mathbf{r}_{ij} \mathbf{F}_{ij}^{N}g(\mathbf{k}) e^{i\mathbf{k} \cdot \mathbf{r}_{ij}} \label{eqn:mol_pressure_tensor}
\end{eqnarray}
where, $\mathbf{F}_{ij}^{N}$ represents the total intermolecular force on molecule $i$ due to molecule $j$. $\mathbf{r}_{ij}=\mathbf{r}_{j}-\mathbf{r}_{i}$ is the minimum image separation of the center of mass of molecule $i$ from the center of mass of molecule $j$. In cases where two sites on two different periodic images of the same molecule may interact, the correct value of $\mathbf{r}_{ij}=\mathbf{r}_{j}-\mathbf{r}_{i}$ corresponding to the particular images of molecule $i$ and $j$ in $\mathbf{F}_{i\alpha j\beta}$ must be used. Though this is unlikely to happen in diatomic fluids it is particularly important in simulation of long molecules. The momenta appearing in these equations, $\mathbf{p}_{i\alpha}$, $\mathbf{p}_{i}$, are the momenta appearing in the respective atomic and molecular equations of motion Eqs.~(\ref{eqn:atomic_hamilton_eqm_isokinetic}) and  (\ref{eqn:mol_hamilton_eqm_isokinetic}).

\subsection{Wave-vector and frequency dependent viscosity \label{sect:IIc}}

The complex wave-vector and frequency dependent viscosity can be evaluated by using two different expressions in terms of the Fourier-Laplace transform of the transverse momentum density autocorrelation function (ACF), $C_{\perp}(\mathbf{k},t)$, and the Fourier-Laplace transform of the stress tensor autocorrelation function, $N(\mathbf{k},t)$ \cite{evans_1990}. We define the complex Laplace transform (one-sided Fourier transform) as $\mathcal{L} [f(t)]=\tilde{f}(w)=\int_{0}^{\infty}f(t)e^{-i\omega t}dt$. We also note that for the sake of simplicity of notation and consistency with the notation used in previous publications, in what follows, we drop the tilde sign over the Fourier transformed correlation functions and keep the tilde notation over the Fourier-Laplace transformed correlation functions only. If we set $\Gamma_{\mathbf{k}}=(0,k_{y},0)$ and $J_{x}$ is the component of the momentum density in the \textit{x} direction, the expression for the wave-vector and frequency dependent viscosity in terms of $\tilde{C}_{\perp}(k_{y},\omega)$ takes the form \cite{evans_1990}:
\begin{equation}
\tilde{\eta}(k_{y},\omega)=\frac{\rho}{k_{y}^{2}}\frac{C_{\perp}(k_{y},t=0)-i\omega \tilde{C}_{\perp}(k_{y},\omega)}{\tilde{C}_{\perp}(k_{y},\omega)} \label{eqn:etakw}
\end{equation}
where $\rho$ is the number density of the fluid and $\tilde{C}_{\perp}(k_{y},\omega)$ is the Laplace transform of the ensemble averaged transverse momentum density autocorrelation function $C_{\perp}(k_{y},t)$, which is defined as
\begin{equation}
C_{\perp}(k_{y},t)=\frac{1}{V}\Big\langle J_{x}(k_{y},t) J_{x}(k_{y},t=0)\Big\rangle . \label{eqn:eqtcacf}
\end{equation}
The zero time value of $C_{\perp}(k_{y},t=0)$ in the thermodynamic limit is
\begin{equation}
C_{\perp}(k_{y},t=0) = \rho k_{B}T . \label{eqn:zteqtcacf}
\end{equation}
$k_{B}$ is Boltzmann's constant. Due to finite time averaging and finite system size, the value of $C(k_y,t=0)$ obtained from simulation differs insubstantially from the exact value given by 
\begin{equation}
C_{\perp}(k_{y},t=0) = \rho k_{B}T \frac{3N-4}{3N} \label{eqn:ezteqtcacf}
\end{equation}
because the total peculiar kinetic energy and three components of the momenta are constants of the motion in our simulations. We also note that the number of degrees of freedom in the simulated system has not been taken into account in Eq.~(\ref{eqn:zteqtcacf}), therefore we use the simulated value in our calculations to ensure numerical consistency of the computed properties. 

The expression for the wave-vector and frequency dependent viscosity in terms of the autocorrelation function of the shear stress $\tilde{N}(k_{y},\omega)$ takes the form:
\begin{equation}
\tilde{\eta}(k_{y},\omega)=\frac{\tilde{N}(k_{y},\omega)}{C(k_{y},t=0)/{\rho}{k_B}T - k^{2} \tilde{N}(k_{y},\omega)/i\omega\rho}
\label{eqn:etakw_N}
\end{equation}
where 
\begin{equation}
\tilde{N}(k_{y},\omega)=\frac{1}{Vk_{B}T}\mathcal{L} \Big[ \Big \langle P_{xy}(k_{y},t) P_{xy}(k_{y},0) \Big\rangle \Big] . \label{eqn:eqnacf}
\end{equation}
In the zero wave-vector limit, a generalization of the Green-Kubo expression for the shear viscosity allows the transverse momentum flux to be in an arbitrary direction rather than along a coordinate axis and can be written in terms of the stress tensor as \cite{hansen_1986,daivis_1994}:
\begin{equation}
\eta=\frac{V}{10k_{B}T}\int_{0}^{\infty} \Big\langle \mathbf{P}^{os}(t):\mathbf{P}^{os}(0) \Big\rangle dt \label{eqn:eta_isotropic_1}
\end{equation}
where the \textit{os} superscript denotes the traceless symmetric part of the stress tensor $\mathbf{P}^{os}(t)=\frac{1}{2}[\mathbf{P}(t)+\mathbf{P}^{T}(t)]-\frac{1}{3}tr[\mathbf{P}(t)]\mathbf{1}$ and $V$ is the simulation volume. In an isotropic fluid, because the tensor, $\mathbf{P}^{os}$, appearing in Eq.~(\ref{eqn:eta_isotropic_1}) is traceless and symmetric, the shear viscosity is also traceless and symmetric. Consequently, the shear viscosity can be expressed in terms of the invariant $I$ of the shear viscosity tensor as $\eta=I/10$.

In the case $\omega \to 0$ and $k_{y} \to 0$, relation ~(\ref{eqn:etakw_N}) reduces to the Green-Kubo formula \cite{hansen_1986}. All the non-zero wave-vector integrals, Eq.~(\ref{eqn:eqnacf}), converge to zero \cite{evans_1990} therefore the relation in Eq.~(\ref{eqn:etakw}) must be used when non-zero wave-vector viscosities are calculated. By computing the integrals one can computationally verify the zero values of the zero-frequency limit of the function $\tilde{N}(\mathbf{k},\omega)$ and thus demonstrate why neither substitution of $\omega=0$ into Eq.(\ref{eqn:etakw_N}) nor evaluation of Eq.~(\ref{eqn:eqnacf}) at non-zero wave-vector yields the zero-frequency wave-vector dependent viscosity.

\section{Simulation details \label{sect:III}}

We use the Edberg, Evans, and Morriss algorithm \cite{edberg_1986,edberg_1987_2,daivis_1992} with an improved cell neighbour list construction algorithm \cite{matin_2003} to perform equilibrium simulations at constant $N$,$V$,$T_{M}$. $N$ is either the number of atoms or molecules and $T_{M}$ is the molecular temperature as defined by Eq.~(\ref{eqn:molecular_temp}). For the atomic fluids studied in this work, the atoms interact via a Lennard-Jones (LJ) or Weeks-Chandler-Andersen (WCA) potential energy function \cite{weeks_1971}. The LJ atoms have an interaction potential truncated at $r_{c}=2.5\sigma$ and WCA atoms have an interaction potential truncated at $r_{c}=2^{1/6}\sigma$ \cite{weeks_1971}. In general:
\begin{equation}
\Phi_{ij}(r_{ij})=
\begin{cases}
    4 \epsilon \Bigg[ { \bigg( \frac{\displaystyle \sigma}{\displaystyle r_{ij}} \bigg) }^{12} -
{\bigg( \frac{\displaystyle \sigma}{\displaystyle r_{ij}} \bigg) }^{6} \Bigg] - \Phi_{c}, & r_{ij} < r_{c} \\
    0,  & r_{ij} \ge r_{c} 
\end{cases}
\label{eqn:WCA_pot}
\end{equation}
where $r_{ij}$ is the interatomic separation, $\epsilon$ is the potential well depth, and $\sigma$ is the value of $r_{ij}$ at which the unshifted potential is zero. The shift $\Phi_{c}$ is the value of the unshifted potential at the cutoff $r_{ij}=r_{c}$, and is introduced to eliminate the discontinuity in the potential energy. At distances greater than the cutoff distance $r_c$, the potential is zero. 

Our diatomic model of liquid chlorine is similar the the one used by Edberg \textit{et al.} \cite{edberg_1987}, Hounkonnou \textit{et al.} \cite{hounkonnou_1992}, Travis \textit{et al.} \cite{travis_1995,travis_1995_a} and more recently by Matin \textit{et al.} \cite{matin_2000, matin_2000_erratum} to allow a direct comparison of our results with previous work. This model represents chlorine as a diatomic LJ molecule with $r_{c}=2.5\sigma$ and a fixed bond length of $0.63\sigma$. For an adequate representation of the properties of chlorine the LJ parameters are: $\sigma=3.332$\AA{} and $\epsilon/k_{B}=178.3$K. Liquid chlorine systems of 108 and 864 molecules are studied at a reduced site number density of $\rho_{a}=1.088$ and a reduced molecular temperature $T_{M}=0.970$. We summarize the most important simulation parameters in Table~\ref{tab:sim_details_table}.
\begin{table}[h]
\caption{Simulation details}
\scalebox{0.8}{
\begin{tabular}{ l c  c c c  c c c  c c c }
\hline\hline
			 							&	& & WCA		& & & LJ		& & & Chlorine	&  \\
\hline
Site number density,$\rho_{a}$	&	& & $0.375, 0.480, 0.840$	& & & $0.840$						& & & $1.088$			& \\
Temperature, 				$T$					&	& & $0.765, 1.0$					& & & $0.765, 1.0$			& & & $0.97$			& \\
Number of atoms, 		$N_a$ 			&	& & $108, 2048, 6912$			& & & $108, 2048, 6912$	& & & $216, 1728$	& \\
Number of sites, 		$N_s$				&	& & $1$										& & & $1$								& & & $2$					& \\
Bond length, 				$l$					&	& & -											& & & -									& & & $0.63$			& \\
LJ cutoff, 					$r_{c}$			&	& & $2^{1/6}$ 						& & & $2.5$ 						& & & $2.5$				& \\
\hline\hline
\end{tabular}}
\label{tab:sim_details_table}
\end{table}
All our simulations were carried out in a cubic box with periodic boundary conditions. The fifth-order Gear predictor corrector algorithm \cite{gear_1966,gear_1971} with time step $\delta t=0.001$ was employed to solve the equations of motion. The equations of motion can be written for a monatomic fluid in the isokinetic ensemble (at equilibrium) as \cite{evans_1990}:
\begin{equation}
\mathbf{\dot{r}}_{i}=\frac{\mathbf{p}_{i}}{m_{i}}, \quad \mathbf{\dot{p}}_{i}=\mathbf{F}_{i} -\zeta_{A}\mathbf{p}_{i}
\label{eqn:atomic_hamilton_eqm_isokinetic}
\end{equation}
where $i$ denotes atom $i$. $\mathbf{r}_{i}$ is the position, $\mathbf{p}_{i}$ is the momentum and $m_{i}$ is the mass of the designated atom. $\mathbf{F}_{i}$ is the force on atom $i$ due to other atoms and $\zeta_{A}$ is the atomic thermostat multiplier. The thermostat multiplier is chosen so as to fix the kinetic temperature. We use the value of $\zeta_{A}$ that results from the application of Gauss' principle of least constraint to the imposition of constant kinetic temperature:
\begin{equation}
\zeta_{A}=\frac{\sum_{i=1}^{N}\mathbf{F}_{i}\cdot \mathbf{p}_{i}}{\sum_{i=1}^{N}\mathbf{p}_{i}^{2}} .
\label{eqn:atomic_thermostat_multiplier}
\end{equation} 
The atomic temperature $T_{A}$ for a system of $N_{a}$ atoms with no internal degrees of freedom is defined as
\begin{equation}
T_{A}= \frac{1}{(d N_{a}-{N_c})k_{B}}\bigg\langle \sum_{i=1}^{N_{a}} \frac{\mathbf{p}_{i}^{2}}{m_{i}} \bigg\rangle
\label{eqn:atomic_temp}
\end{equation}
Here angled brackets denote an ensemble average, $d$ is the dimensionality of the atomic system, $N_c$ is the number of constraints on the system (including constraints for conserved quantities). 

The equations of motion (EOM) for a molecular fluid can be formulated in either atomic or molecular versions. In fact the molecular versions of the homogeneous isothermal EOM with a molecular thermostat at equilibrium are similar to atomic EOM with a molecular thermostat, provided that all of the relevant forces are included \cite{todd_2007}. The thermostatted EOM for molecular systems are given by \cite{todd_2007}
\begin{equation}
\mathbf{\dot{r}}_{i\alpha}=\frac{\mathbf{p}_{i\alpha}}{m_{i\alpha}}, \quad
\mathbf{\dot{p}}_{i\alpha}=\mathbf{F}_{i\alpha} + \mathbf{F}_{i\alpha}^{C} 
- \zeta_{M}\frac{m_{i\alpha}}{M_{i}}\mathbf{p}_{i}
\label{eqn:mol_hamilton_eqm_isokinetic}
\end{equation}
where $i\alpha$ denotes site $\alpha$ on molecule $i$. $\mathbf{r}_{i\alpha}$ is the position, $\mathbf{p}_{i\alpha}$ is the momentum and $m_{i\alpha}$ is the mass of the site. The force on a site is separated into two terms: $\mathbf{F}_{i\alpha}$ is the contribution due to the Lennard-Jones type interactions on site $\alpha$ of molecule $i$ and $\mathbf{F}_{i\alpha}^{C}$ is either the constraint force or the bonding force. $\zeta_{M}$ is the molecular thermostat multiplier, given by 
\begin{equation}
\zeta_{M}=\frac{\sum_{i=1}^{N_{m}}\mathbf{F}_{i}\cdot \mathbf{p}_{i}/M_{i}}{\sum_{i=1}^{N_{M}}\mathbf{p}_{i}^{2}/M_{i}}
\label{eqn:mol_thermostat_multiplier_2}
\end{equation}
where $\mathbf{F}_{i}$ is the total force acting on molecule $i$ and $M_{i}$ is the mass of molecule $i$.
$\zeta_{M}$ is derived from Gauss' principle of least constraint and acts to keep the molecular center of mass kinetic temperature $T_M$ constant. Several algorithms are available for this purpose \cite{allen_1989}. The molecular temperature $T_M$ is defined by
\begin{equation}
T_{M}= \frac{1}{(d N_{m}-N_{c})k_{B}}\bigg\langle \sum_{i=1}^{N} \frac{\mathbf{p}_{i}^{2}}{m_{i}} \bigg\rangle \label{eqn:molecular_temp}
\end{equation}
where $N_c$ is the number of translational center-of-mass degrees of freedom and depends on the total number of sites and the number of constraints on the system. The details of the constraint algorithm used to calculate $\mathbf{F}_{i\alpha}^{C}$ have been discussed previously \cite{edberg_1986,edberg_1987,morriss_1991}.
All the systems were equilibrated for at least $10^6$ time steps. The results from production runs were ensemble averaged over 14 runs, each of length $10^6$ steps (i.e. a total of $1.4 \times 10^7$ time steps). The transverse momentum density ACFs were computed over at least $20$ reduced time units  and the stress ACFs were computed over at least $40$ reduced time units. Both the transverse momentum density and stress ACFs were computed at wave-vectors $k_{yn}=2\pi n/L_{y}$ where mode number $n$ is from 0 to 40 with increment 2 and $L_{y}=[N_{a}/ \rho]^{1/3}$. For the remainder of this paper we drop the $n$ index in $k_{yn}$ for simplicity. The ACFs were Laplace transformed with respect to time using Filon's rule \cite{allen_1989}. We do not report the frequency dependent viscosities in this work. The wave-vector and frequency dependent viscosities were calculated using Eqs.~(\ref{eqn:etakw}) and~(\ref{eqn:etakw_N}). Eq.~(\ref{eqn:etakw}) was used to obtain the non-zero wave-vector and frequency dependent viscosities and Eq.~(\ref{eqn:etakw_N}) was used to obtain the zero wave-vector viscosity. 

In this work, all quantities are expressed in reduced Lennard-Jones units. Thus, our reference units are the: reduced length $r^{*}=r/ \sigma$, reduced number density $\rho^{*}=\rho / \sigma^{3}$, reduced temperature $T^{*}=k_{B} T/ \epsilon$, reduced time $t^{*}=t/{(\sigma (m/ \epsilon)}^{1/2}$, reduced pressure $\mathbf{P}^{*}=\mathbf{P}(\sigma^{3}/\epsilon)$, reduced energy $E^{*}=E/\epsilon$ and reduced viscosity $\eta^{*}=\eta\sigma^{2}/ \sqrt{(m\epsilon)}$. For the remainder of this paper we apply these units and omit writing the asterisk. We will not distinguish between $T_{M}$ and $T_{A}$, but simply use $T$ to indicate the temperature.

\section{Results and discussion\label{sect:IV}}
The autocorrelation functions were evaluated for both 108 and 864 molecule system in order to determine whether the results were system size dependent. There were no observed differences within the statistical errors in both monatomic and diatomic systems. We also note that in order to limit the number of figures, we do not display the results for the transverse momentum density and stress autocorrelation functions. However, we must mention that for monatomic systems both quantities (i.e. the transverse momentum density and stress ACFs) were in good agreement with those previously observed for Lennard-Jones monatomic liquids and their running integrals have fully converged which suggests that the correlation functions have decayed to zero. 

\subsection{Viscosity kernels in reciprocal space \label{sect:IVb}}
\begin{figure}
\begin{center}
\includegraphics[scale=0.75]{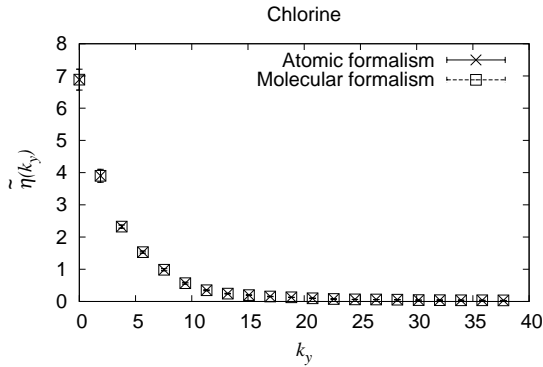}
\caption{\label{fig:etak_am} $\tilde{\eta}(k_{y})$ versus $k_{y}$ for chlorine calculated using atomic and molecular formalisms ($\rho_{a}=1.088$, $T=0.97$, $N_{a}=1728$).}
\end{center}
\end{figure}

\begin{figure*}
\begin{center}
\includegraphics[scale=1.0]{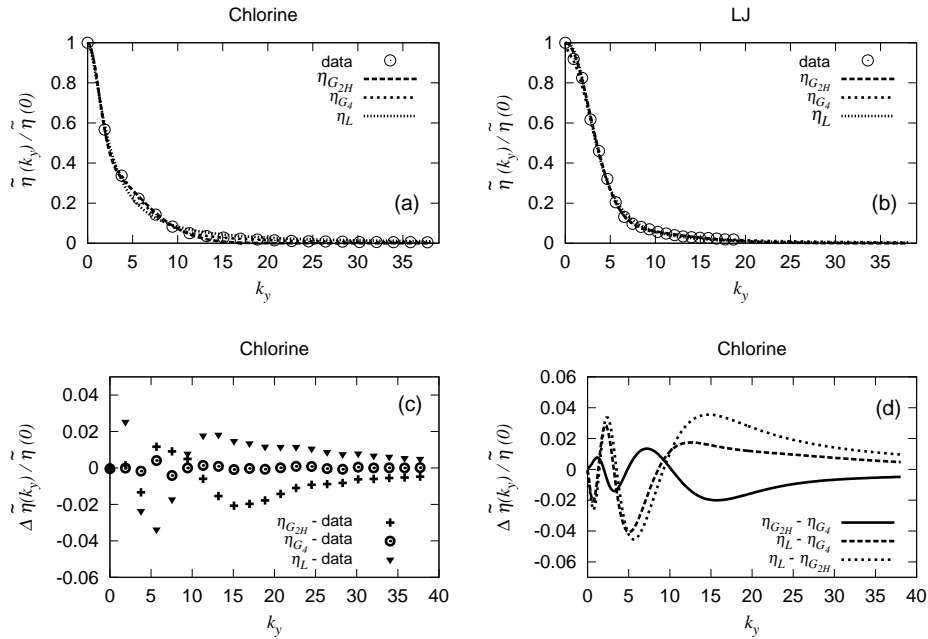}
\caption{\label{fig:etak_1} Normalized kernel data, best fit of Eq.~(\ref{eqn:fit_gauss}) with $N_{G}=2$ and $4$, and Eq.~(\ref{eqn:fit_lorentz}) and difference between the fits:
(a) Best fits to normalized kernel for chlorine fluid ($\rho_{a}=1.088$, $T=0.97$, $N_{a}=1728$);   
(b) Best fits to normalized kernel for LJ fluid ($\rho_{a}=0.840$, $T=1.0$, $N_{a}=2048$);   
(c) Differences between the kernels and simulation data (a); 
(d) Differences between the fitted kernels (b).}
\end{center}
\end{figure*}

\begin{figure*}
\begin{center}
\includegraphics[scale=1.0]{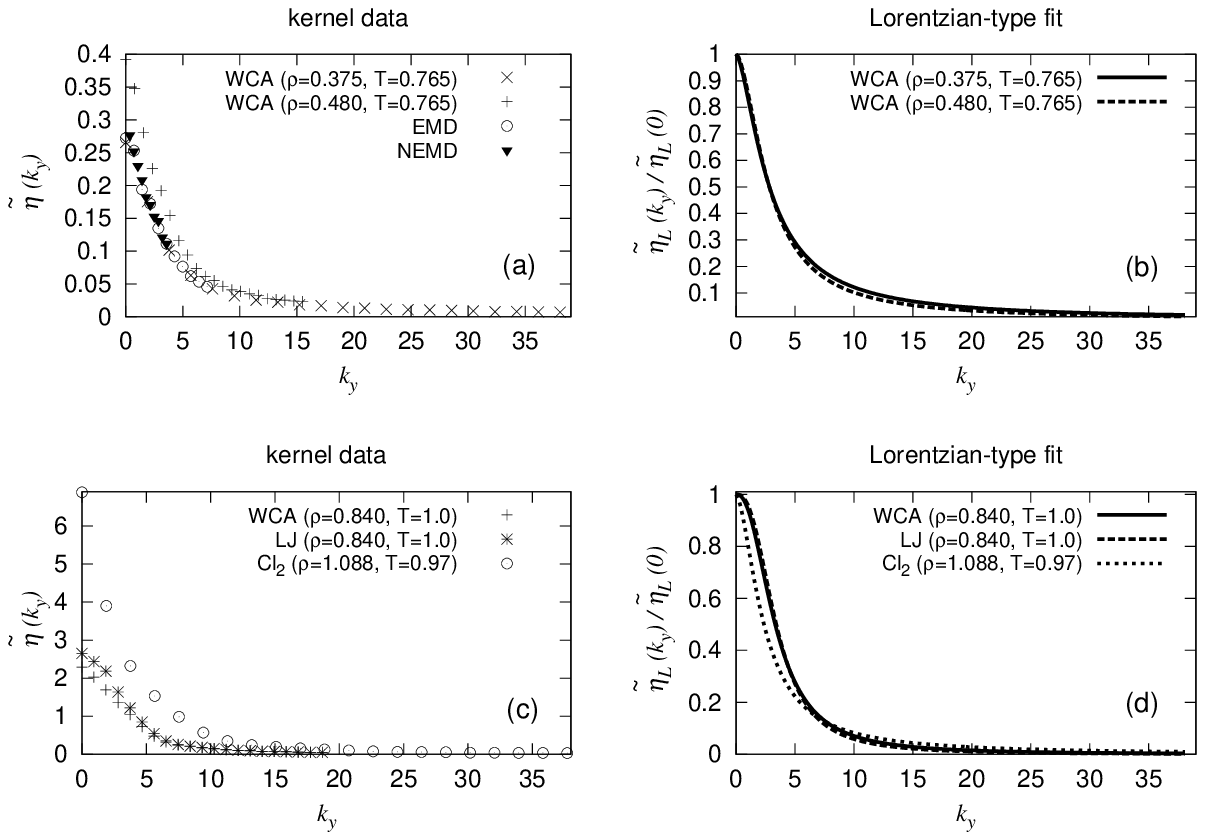}
\caption{\label{fig:etak_2} Reciprocal-space kernels of monatomic and diatomic fluids:
(a) Kernel data of a WCA fluid at two different densities ($T=0.765$, $N_{a}=2048$); 
(b) Best fit of normalized kernel data (a) to Lorentzian-type function Eq.~(\ref{eqn:fit_lorentz}); 
(c) Kernel data of a WCA fluid and LJ fluid at the same state point ($\rho_{a}=0.840$, $T=1.0$, $N_{a}=2048$), Chlorine at ($\rho_{a}=1.088$, $T=0.97$, $N_{a}=1728$);   
(d) Best fit of the normalized kernel data (c) to the Lorentzian-type function Eq.~(\ref{eqn:fit_lorentz}).}
\end{center}
\end{figure*}

The reciprocal space kernels for monatomic and diatomic fluids are plotted in figures~(\ref{fig:etak_am}-\ref{fig:etak_2}). The error bars are smaller than the symbol sizes and therefore omitted in figures~\ref{fig:etak_1} and \ref{fig:etak_2}. Generally the statistical reliability of reciprocal space kernel data increases as $k_{y}$ increases.
\begin{table*}
\caption{Zero frequency, zero wave-vector shear viscosity and fitted parameter values for monatomic and diatomic systems}
\scalebox{0.9}{
\begin{tabular}{ l l  l  c c c c c c c c c c c c c c c c }
\hline\hline
&	& & WCA	& WCA & WCA & LJ	& LJ & \multicolumn {1}{c}{Chlorine} \\
\textbf{State Point} 						
&	& $\rho_{a}$	
& \multicolumn {1}{c}{$0.375$}	
& \multicolumn {1}{c}{$0.480$} 	
& \multicolumn {1}{c}{$0.840$}		
& $0.840$ & $0.840$	& $1.088$ \\
&	& $T$ 		
& \multicolumn {1}{c}{$0.765$}	
& \multicolumn {1}{c}{$0.765$}	
& \multicolumn {1}{c}{$1.000$}		
& $0.765$ & $1.0$		& $0.97$ 	\\
\textbf{System size} 						
&	& $N_a$		& \multicolumn {5}{c}{$2048$}		& $1728$	\\
\hline
&	& ${\eta}_{0}$	& $0.265(0.273$\cite{hansen_2007_2})   & $0.392$ & $2.290$ & $2.810$ & $2.650$  & $6.889$	\\
\hline
\textbf{2-term Gaussian}, $A_{2}=1-A_{1}$ Eq.~(\ref{eqn:fit_gauss})		
&	& $A$					& $0.189(0.440$\cite{hansen_2007_2}) & $0.309$ & $0.155$ & $0.093$ & $0.107$ & $0.407$\\
& & $\sigma_1$	& $12.48(4.750$\cite{hansen_2007_2}) & $6.916$ & $8.122$ & $10.04$ & $9.088$ & $5.377$\\
&	& $\sigma_2$  & $2.116(1.376$\cite{hansen_2007_2}) & $1.835$ & $2.592$ & $2.778$ & $2.759$ & $1.236$\\ 
&	& $s_{r}$	  	& $0.007(0.005$\cite{hansen_2007_2}) & $0.013$ & $0.044$ & $0.021$ & $0.027$ & $0.082$	\\ 
\hline
\textbf{2-term Gaussian} Eq.~(\ref{eqn:fit_gauss})	
&	& $A_1$				& $0.792$  	& $0.687$ 	& $0.874$ & $0.907$ 		& $0.892$  	& $0.592$		\\
&	& $A_2$				& $0.174$  	& $0.254$ 	& $0.155$ & $0.094$ 		& $0.106$  	& $0.407$		\\
&	& $\sigma_1$	& $2.245$ 	& $2.113$  	& $2.592$ & $2.776$ 		& $2.765$ 	& $1.237$		\\
&	& $\sigma_2$  & $13.36$  	& $7.745$ 	& $8.124$ & $10.02$ 		& $9.127$ 	& $5.377$		\\ 
&	& $s_{r}$  		& $0.007$  	& $0.011$ 	& $0.035$ & $0.022$ 		& $0.031$ 	& $0.081$		\\ 
\hline
\textbf{4-term Gaussian} Eq.~(\ref{eqn:fit_gauss})	
&	& $A_1$				& $0.432$  	& $0.566$		& $0.778$ & $0.689$ 		& $0.868$ 	& $0.398$		\\
&	& $A_2$				& $0.394$ 	& $0.248$ 	& $0.118$ & $0.190$ 		& $0.047$ 	& $0.538$		\\
&	& $A_3$				& $0.120$  	& $0.138$ 	& $0.088$ & $0.114$ 		& $0.089$ 	& $0.055$		\\
&	& $A_4$				& $0.056$  	& $0.047$ 	& $0.017$ & $0.017$ 		& $0.020$ 	& $0.008$		\\
&	& $\sigma_1$ 	& $3.228$ 	& $2.826$ 	& $2.950$ & $2.709$ 		& $2.814$ 	& $4.355$		\\
&	& $\sigma_2$  & $1.261$ 	& $0.821$ 	& $0.651$ & $2.709$ 		& $0.145$ 	& $1.155$		\\
&	& $\sigma_3$ 	& $8.165$ 	& $6.973$ 	& $8.496$ & $7.037$ 		& $7.628$ 	& $10.46$		\\
&	& $\sigma_4$  & $15.19$ 	& $14.74$ 	& $23.66$ & $24.94$ 		& $19.99$ 	& $37.56$		\\ 
&	& $s_{r}$  		& $0.008$  	& $0.002$ 	& $0.012$ & $0.014$ 		& $0.024$ 	& $0.018$		\\ 
\hline
\textbf{Lorentzian-type} Eq.~(\ref{eqn:fit_lorentz})	
&	& $\alpha$ 	& $0.198(0.180$\cite{hansen_2007_2}) & $0.170$ & $0.062$ & $0.041$ & $0.043$ & $0.239$	\\
&	& $\beta$		& $1.562(1.662$\cite{hansen_2007_2}) & $1.715$ & $2.326$ & $2.602$ & $2.572$ & $1.667$	\\ 
&	& $s_{r}$	 	& $0.002(0.005$\cite{hansen_2007_2}) & $0.005$ & $0.042$ & $0.018$ & $0.042$ & $0.016$	\\ 
\hline\hline
\end{tabular}}
\label{tab:atomic_param}
\end{table*} 
Our zero wave-vector, zero frequency viscosities for monatomic fluids agree well with those available in the literature. For the WCA system at the state point ($\rho_{a}=0.375$, $T=0.765$) we found 
${\eta}_{0}=0.27{\pm}0.01$ which agrees with the results of Hansen \textit{et al.} $0.273$ \cite{hansen_2007_2,alley_1983}, while at the state point ($\rho_{a}=0.840$, $T=1.0$) we found ${\eta}_{0}=2.29{\pm}0.07$, in agreement with the results of Matin \textit{et al.} ($2.1\pm0.2$) 
\cite{matin_2000}. For chlorine we found ${\eta}_{0}=6.89{\pm}0.32$, which agrees with the limiting values (6.7${\pm}$0.4) of the shear or elongational viscosities at zero strain rate \cite{matin_2000}.

The wave-vector dependent viscosity for diatomic systems, figure~\ref{fig:etak_am}, depicts a similar behaviour within the statistical uncertainty in both atomic and molecular formalisms. 

It has been shown previously that numerous one parameter functions failed to capture the behaviour of the reciprocal space kernel data \cite{hansen_2007_2}. We therefore present the best fits with two or more fitting parameters. We have identified two functional forms that fit the data well: an $N_{G}$ term Gaussian function
\begin{equation}
\tilde{\eta}_{G}(k_{y})={\eta}_{0} \sum_{j}^{N_{G}} A_{j} \exp(-k^{2}_{y}/2\sigma^{2}_{j}) \qquad A_{j}, \sigma_{j} \in {\mathbb{R}_{+}}
\label{eqn:fit_gauss} 
\end{equation}
and a Lorentzian type function
\begin{equation}
\tilde{\eta}_{L}(k_{y})=\frac{{\eta}_{0}}{1+\alpha \left| k_{y} \right|^\beta } \qquad \alpha , \beta \in {\mathbb{R}_{+}} ,
\label{eqn:fit_lorentz} 
\end{equation}
We present the best fits of the data to (i) a two-term Gaussian function with freely estimated amplitudes (i.e. unconstrained fitting) termed as $\tilde{\eta}_{G_{2}}$, (ii) to a two-term Gaussian function with interdependent amplitudes (i.e. constrained fitting $\sum_{j}^{N_{G}} A_{j}=1$) given by Hansen \textit{et al.} \cite{hansen_2007_2} and termed as $\tilde{\eta}_{G_{2H}}$, (iii) to a four-term Gaussian function with freely estimated amplitudes, termed as $\tilde{\eta}_{G_{4}}$ and (iv) to the Lorentzian type function, Eq.~(\ref{eqn:fit_lorentz}). In order to measure the magnitude of the residuals we use the residual standard deviation defined as $s_{r}=\sqrt{\sum_{n=1}^{n_{s}}r^{2}/(n_{s}-n_{p})}$ where $n_s$ is the number of data points, $n_p$ is the number of fitting parameters, and $r$ is the residual \cite{peck_2008}. After an iterative curve fitting procedure the accurate estimation of ${\eta}_0$ was kept fixed allowing all other parameters in Eqs.~(\ref{eqn:fit_gauss}) and ~(\ref{eqn:fit_lorentz}) to be used as fitting parameters. In Table~\ref{tab:atomic_param} we have listed the fitting parameters for monatomic and diatomic molecular fluids and compared to the previous results where possible. 

A useful check of the fitting can be performed by calculating the total Gaussian amplitudes which should converge to the value of 1, Table~\ref{tab:tot_amp}. 

\begin{table}
\caption{Total amplitude for Gaussian functional form}
\scalebox{0.8}{
\begin{tabular}{ l l  l  c c c c c c c c c c c c c c c c }
\hline\hline
&	& & \multicolumn {3}{c}{WCA}	& \multicolumn {2}{c}{LJ}		& \multicolumn {1}{c}{Chlorine}\\
\textbf{State Point} 						
&	& $\rho_{a}$		
& \multicolumn {1}{c}{$0.375$}	
& \multicolumn {1}{c}{$0.480$} 	
& \multicolumn {1}{c}{$0.840$}		& $0.840$		& $0.840$		& $1.088$ \\
&	& $T$ 			
& \multicolumn {1}{c}{$0.765$}	
& \multicolumn {1}{c}{$0.765$}	
& \multicolumn {1}{c}{$1.000$}		& $0.765$ 	& $1.0$			& $0.97$ \\
\hline
\textbf{2-term Gaussian} 	
&	& $\sum_{j=1}^{2}{A}$	& $0.966$  		& $0.941$ 		& $1.029$ 	& $1.001$ 	& $0.998$  	& $0.999$	\\
\textbf{4-term Gaussian} 	
&	& $\sum_{j=1}^{4}{A}$	& $1.002$  		& $0.999$			& $1.001$ 	& $1.010$ 	& $1.024$ 	& $0.999$	\\
\hline\hline
\end{tabular}}
\label{tab:tot_amp}
\end{table}

The reciprocal space results presented in figure~\ref{fig:etak_1} for LJ and chlorine systems demonstrate that the four-term Gaussian function fits the data much better than the other two forms with a difference 
between the data and the fit of less than 0.5$\%$, see figure~\ref{fig:etak_1}(c). The two-term Gaussian $\tilde{\eta}_{G_{2H}}$ fits the kernel data better than the Lorentzian-type function in the low-$k_{y}$ region, figure~\ref{fig:etak_1}(a), which suggests a more Gaussian-like behaviour in the low-$k_{y}$ region, a fact previously observed by Hansen \textit{et. al} \cite{hansen_2007_2} for atomic fluids modeled with WCA potentials. Nevertheless the difference between the two-term Gaussian fit and data is less than 2$\%$ which still makes the $\tilde{\eta}_{G_{2H}}$ a good analytical three parameter approximation of the reciprocal space viscosity kernel. The maximum difference between the Lorentzian-type fit and Gaussian fits are around 4$\%$ while the maximum difference between the Gaussian fits is about 2$\%$, see figure~\ref{fig:etak_1}(d). Essentially, this suggests that, when computing the real space kernels, the four-term Gaussian functional form is to be trusted. It is obvious that eight parameters in the four-term Gaussian make its use less convenient, but on the other hand the Gaussian function can analytically be inverse Fourier transformed while the inverse Fourier transform of the Lorentzian-type function can only be evaluated numerically for general values of $\beta$. 

Figure~\ref{fig:etak_2}(a) shows the kernel data for a WCA fluid at two different densities along with two sets of data published previously by Hansen \textit{et al.} \cite{hansen_2007_2}. EMD is the set obtained from an equilibrium MD simulation at the same state point ($\rho_{a}=0.375$, $T=0.765$) and NEMD is the set obtained from a nonequilibrium MD simulation based on the sinusoidal transverse force (STF) method. An excellent agreement between both sets of data was found. Figure~\ref{fig:etak_2}(b) shows the normalized fit to Eq.~(\ref{eqn:fit_lorentz}). The normalized kernels, figure~\ref{fig:etak_2}(b), show a similar behavior for $k_n \le 4$. Though the higher density kernel is slightly lower for $k_n \ge 4$ they show a similar limiting behavior. This effect was not seen by Hansen \textit{et al.} due to lack of data for high wave vectors. Figure~\ref{fig:etak_2}(d) indicates that despite the difference between the interaction potentials, the results for LJ and WCA fluids are very close. This confirms that transport is dominated by repulsive interactions, rather than attractive. The sharper kernel for the diatomic system, figure~\ref{fig:etak_2}(d), suggests a more Lorentzian-type behavior in the low wave-vector region. It is also important to mention that even though the fitting parameters are significantly affected by temperature, the resulting kernels vary weakly over the range of temperatures chosen here. This was also observed by Hansen \textit{et al.} for WCA monatomic fluids. 

\subsection{Viscosity kernels in physical space \label{sect:IVc}}

\begin{figure*}
\begin{center}
\includegraphics[scale=1.0]{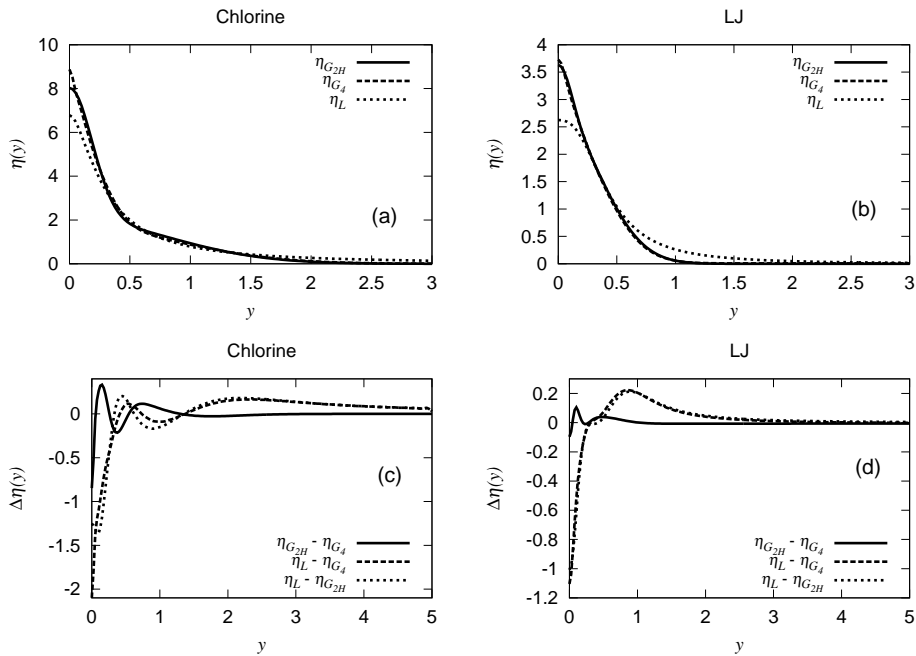}
\caption{\label{fig:etay_1} Real space kernel of monatomic and diatomic fluids as predicted by Gaussian and Lorentzian fits of reciprocal space kernel data:
(a) chlorine ($\rho_{a}=1.088$, $T=0.97$);   
(b) LJ ($\rho_{a}=0.840$, $T=1.0$);   
(c) Differences between the kernels shown in (a); 
(d) Differences between the kernels shown in (b).}
\end{center}
\end{figure*}
The viscosity kernel in reciprocal space is an even function since it is symmetric about the origin; thus the the real space kernel is symmetric because the Fourier transform keeps the even properties of the function. This means that the viscosity kernel in physical space can be found via an inverse Fourier cosine transform, $F_{c}^{-1}[\dots]$ (a special case of the continuous Fourier transform arising naturally when attempting to transform an even function), of the viscosity kernel in reciprocal space. Since the integral is being computed over an interval symmetric about the origin (i.e. -$\infty$ to +$\infty$), the second integral must vanish to zero, and the first may be simplified to give:
\begin{equation}
F_{c}^{-1}[\tilde{\eta}(k_{y})]=\eta (y)=
\sqrt{\frac{2}{\pi}}\int\limits_{0}^{\infty}\tilde{\eta} (k_{y})\cos(k_{y}y)dk_y . \label{eqn:ifct} 
\end{equation}
The inverse Fourier cosine transform of the Gaussian function, Eq.~(\ref{eqn:fit_gauss}), exists \cite{papoulis_1962} and it is even possible to obtain an analytical expression. For an $N_{G}$ term Gaussian function the inverse Fourier cosine transform is
\begin{equation}
\eta_{G}(y)=\frac{{\eta}_{0}}{\sqrt{2\pi}} \sum_{j}^{N_{G}} A_{j} \sigma_{j} \exp[-(\sigma_{j}y)^{2}/2] \qquad A_{j}, \sigma_{j} \in {\mathbb{R}_{+}} . \label{eqn:ft_fit_gauss} 
\end{equation}

Though the Lorentzian-type function given in Eq.~(\ref{eqn:fit_lorentz}) fulfills the criteria for having an inverse Fourier transform $\eta_{L}(y)$ (i.e. the function is absolutely integrable, square integrable and the function and its derivative are piecewise continuous), the integral in Eq.~(\ref{eqn:ifct}) is not readily obtained analytically in the general case. However, the integral can be evaluated numerically. In this work, a Simpson method has been employed for this purpose.  

\begin{figure*}
\begin{center}
\includegraphics[scale=1.0]{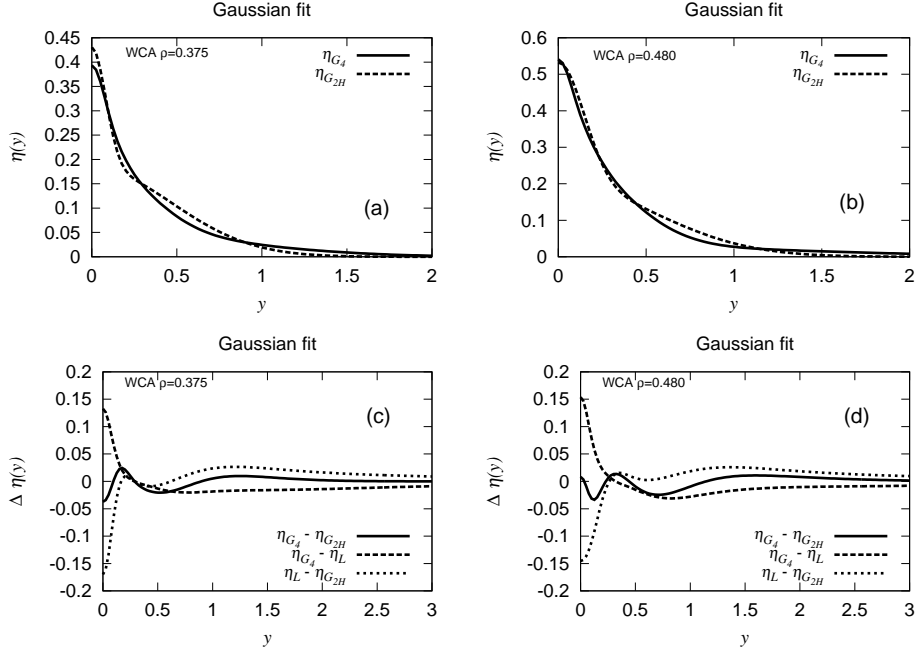}
\caption{\label{fig:etay_3} Real space kernel of monatomic WCA fluids as predicted by Gaussian fits of the reciprocal space kernel data, Eq.~(\ref{eqn:ft_fit_gauss}):
(a) Kernels obtained from two- and four-term Gaussian fits for a WCA fluid at $\rho_{a}=0.375$ and $T=0.765$;   
(b) Kernels obtained from two- and four-term Gaussian fits for a WCA fluid at $\rho_{a}=0.480$ and $T=0.765$;   
(c) Differences between the kernels for a WCA fluid at $\rho_{a}=0.375$ and $T=0.765$; 
(d) Differences between the kernels for a WCA fluid at $\rho_{a}=0.480$ and $T=0.765$.}
\end{center}
\end{figure*}

\begin{figure*}
\begin{center}
\begin{tabular}{cc}
\includegraphics[scale=0.5]{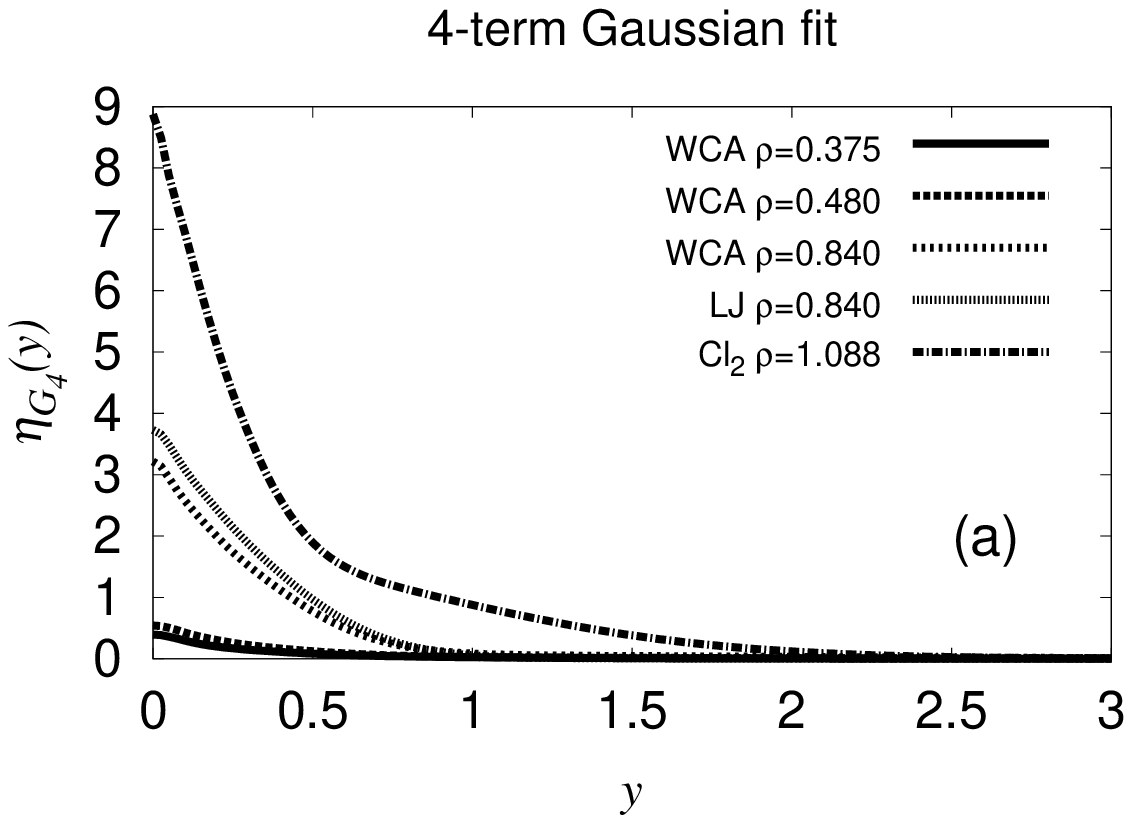} & \includegraphics[scale=0.5]{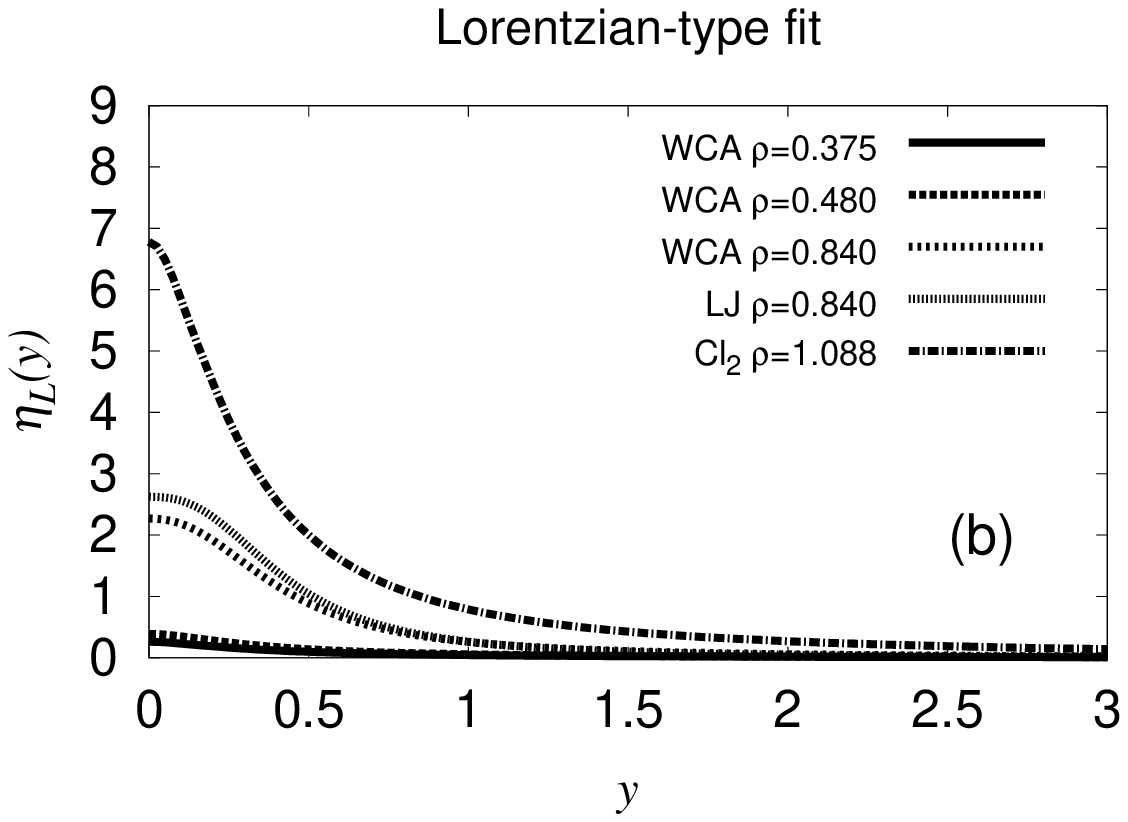}\\
\end{tabular}
\caption{\label{fig:etay_2_1} Real space viscosity kernels of monatomic and diatomic fluids, WCA ($\rho_{a}=0.375$, $T=0.765$), WCA ($\rho_{a}=0.480$, $T=0.765$), WCA ($\rho_{a}=0.840$, $T=1.0$), LJ ($\rho_{a}=0.840$, $T=1.0$), chlorine ($\rho_{a}=1.088$, $T=0.97$): 
(a) Kernels obtained from the four-term Gaussian functional form Eq.~(\ref{eqn:ft_fit_gauss});  
(b) Kernels obtained numerically from the Lorentzian-type functional form Eq.~(\ref{eqn:fit_lorentz}).}
\end{center}
\end{figure*}

The real space kernels for atomic and diatomic fluids at zero frequency are presented in figure~\ref{fig:etay_1}, \ref{fig:etay_3} and \ref{fig:etay_2_1}. Figure~\ref{fig:etay_1}(a) shows the resulting kernels for chlorine and  figure~\ref{fig:etay_1}(b) shows the resulting kernels for a LJ fluid extracted from two-term and four-term Gaussian functions Eq.~(\ref{eqn:ft_fit_gauss}), and inverse Fourier transform of Eq.~(\ref{eqn:fit_lorentz}). We find very little difference between the kernels obtained via two- and four-term Gaussians for these systems. Figures~\ref{fig:etay_1}(c) and \ref{fig:etay_1}(d) show the differences between all three fits. It can be seen that there exists a significant difference (almost 25$\%$) between the kernels extracted from Gaussian and Lorentzian type functions for small $y$. The discrepancy decreases rapidly as $y$ increases and becomes approximately zero for $y \ge 1.5$. The width of the  kernel for chlorine is roughly 4-6 atomic diameters, figure~\ref{fig:etay_1}(a), and 3-5 atomic diameters for monatomic LJ and WCA fluids, figure~\ref{fig:etay_1}(b).

For monatomic systems at relatively low densities ($\rho_{a}=0.375-0.480$), the real space kernels are  affected considerably by the functional form chosen to fit the reciprocal space kernel, figure~\ref{fig:etay_3}. For instance, the equally weighted two-term Gaussian function, figure~\ref{fig:etay_3}(a), distorts the real space kernels and predicts a noticeably higher $\eta(y=0)$ value. As we increase the density, the discrepancy between Gaussian functions, as well as between Gaussian and Lorentzian-type functions, only partially reduces, figure~\ref{fig:etay_3}(b,d). The width of the kernel for WCA fluids at low density is roughly 2-4 atomic diameters.

In figures~\ref{fig:etay_2_1}(a) and \ref{fig:etay_2_1}(b) we compare the unnormalized kernel data in $y$ space extracted from four-term Gaussian and Lorentzian-type functional forms for all the simulated systems. Despite the fact that the difference between the reciprocal kernels is less than 4$\%$, figure~\ref{fig:etak_1}(d) (e.g. chlorine - dashed line), the kernels for the corresponding systems in real space look noticeably different (figure~\ref{fig:etay_2_1}(a) and figure~\ref{fig:etay_2_1}(b)), for all the systems (e.g. chlorine - dashed dotted line). However, the zero wave-vector viscosities obtained from both functional forms are very close, with less than 2$\%$ error. 

\begin{figure}
  \begin{center}
    \includegraphics[scale=0.9]{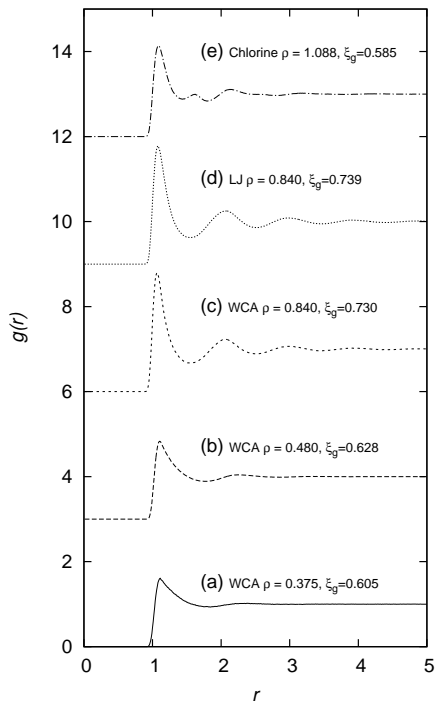}
    \caption{\label{fig:g_r_am} Pair-distance correlation function $g(r)$ and normalization factors $\xi_{g}$, Eq.~(\ref{eqn:norm_g_r}): WCA [(a) $\rho_{a}=0.375$, (b) $\rho_{a}=0.480$ both at $T=0.765$], (c) WCA ($\rho_{a}=0.840$, $T=1.0$); LJ (d) ($\rho_{a}=0.840$, $T=1.0$); chlorine (e) ($\rho_{a}=1.088$, $T=0.97$). For clarity, the RDFs are shifted upwards by 3 units.}
  \label{fig:g_r_ad}
  \end{center}
\end{figure}
\begin{figure*}
\begin{center}
\begin{tabular}{cc}
\includegraphics[scale=0.5]{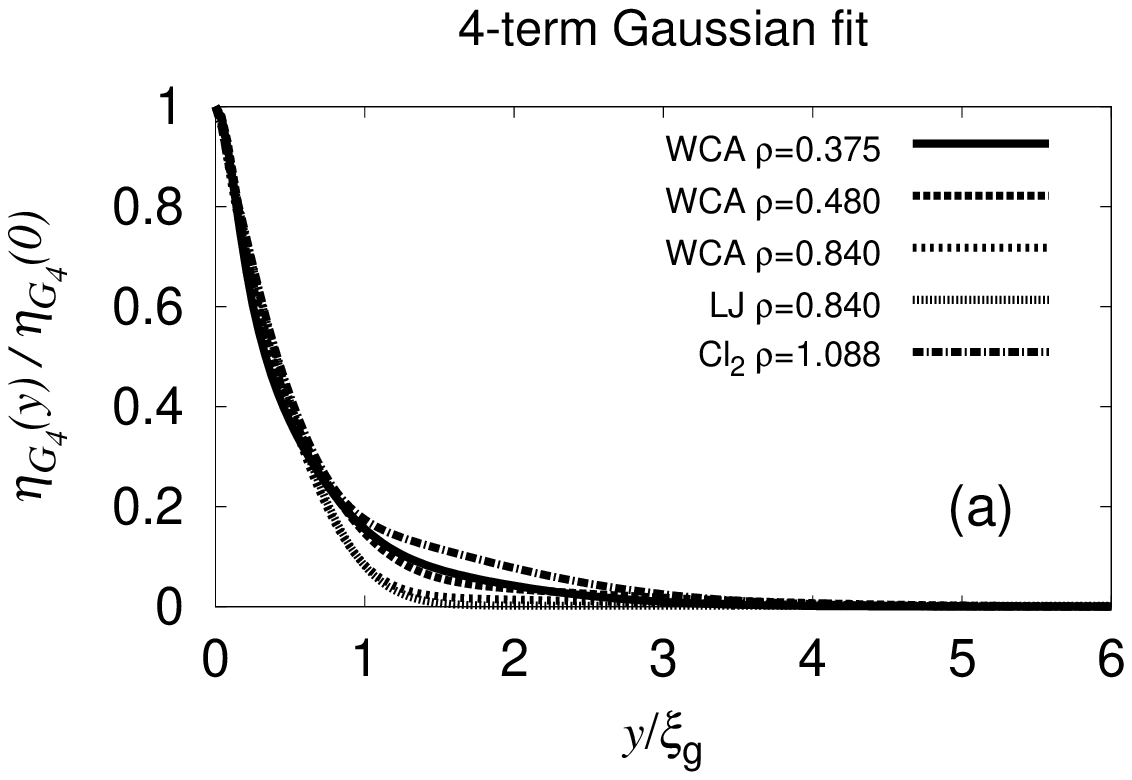} & \includegraphics[scale=0.5]{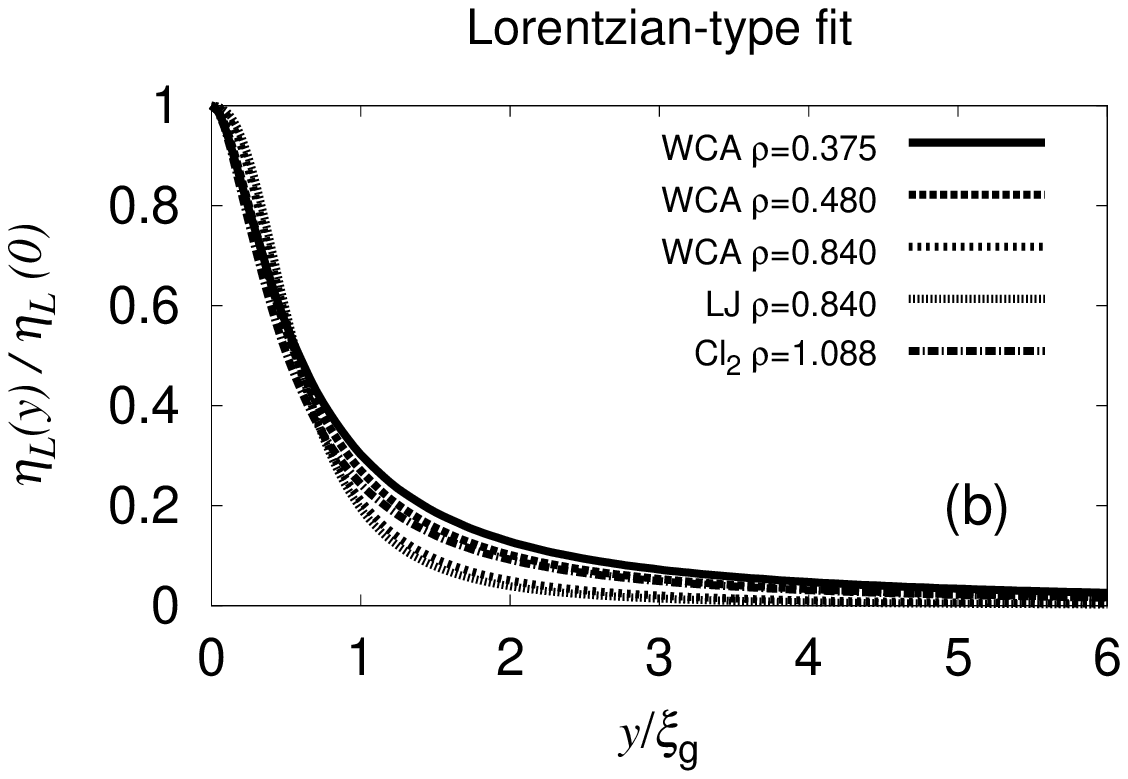}\\
\end{tabular}
\caption{\label{fig:etay_2_2} Normalized real space viscosity kernels of monatomic and diatomic fluids, WCA ($\rho_{a}=0.375$, $T=0.765$), WCA ($\rho_{a}=0.480$, $T=0.765$), WCA ($\rho_{a}=0.840$, $T=1.0$), LJ ($\rho_{a}=0.840$, $T=1.0$), chlorine ($\rho_{a}=1.088$, $T=0.97$): 
(a) Normalized kernels, shown in Fig.~\ref{fig:etay_2_1}(a), obtained from the four-term Gaussian functional form Eq.~(\ref{eqn:ft_fit_gauss});  
(b) Normalized kernels, shown in Fig.~\ref{fig:etay_2_1}(b), obtained numerically from the Lorentzian-type functional form Eq.~(\ref{eqn:fit_lorentz}).}
\end{center}
\end{figure*}

We can determine the zero wave-vector viscosities $\eta_{0}=\eta(k=0,\omega =0)$ by integrating the real space kernel over $y$, and thus test our numerical analysis. The zero wave-vector viscosity $\eta_{0}$ obtained by a Gaussian function $\eta_{G}(y)$ is 
\begin{equation}
\eta_{0}=\int\limits_{-\infty}^{\infty} \eta_{G}(y)dy=\frac{{\eta}_{0}}{\sqrt{2\pi}} 
\int\limits_{-\infty}^{\infty} \Big\{ \sum_{j} A_{j} \sigma_{j} \exp [-(\sigma_{j}y)^{2}/2] \Big\} dy . \label{eqn:lev} 
\end{equation}
Since the general analytical expression for $\eta_{L}(y)$ does not exist \cite{hansen_2007_2} we evaluate the integral numerically and present the results from all functional forms in Table~\ref{tab:am_leffv}. A comparison of the viscosities in Table~\ref{tab:am_leffv} with the simulated zero frequency zero wave-vector shear viscosities given in Table~\ref{tab:atomic_param} shows an integration error of less than 3$\%$. This confirms the accuracy of our numerical analysis techniques. 
\begin{table}
\caption{Effective viscosities evaluated from $\eta_{G_{4}}(y)$, $\eta_{G_{2H}}(y)$ and numerically from $\tilde{\eta}_{L}(k)$. The values can be compared to the zero frequency, zero wave-vector viscosities shown in Table \ref{tab:atomic_param}.}
\scalebox{0.9}{
\begin{tabular}{ l l l c c c c c c c c c c c c c c c c }
\hline\hline
&	& 						& \multicolumn{3}{c}{WCA} 			& \multicolumn{2}{c}{LJ} 	& Chlorine	\\
\textbf{State Point} 							
&	& $\rho_{a}$	& $0.375$ & $0.480$	& $0.840$		& $0.840$	& $0.840$		& $1.088$	\\
&	& $T$ 				& $0.765$	& $0.765$	& $1.000$		& $0.765$	& $1.000$		& $0.97$	\\
\hline
\textbf{2-term Gaussian}	
&	& $\eta_{0}$	& $0.265$	& $0.392$	& $2.290$		& $2.723$	& $2.614$		& $6.881$	\\
\textbf{4-term Gaussian} 	
&	& $\eta_{0}$	& $0.265$	& $0.390$	& $2.288$		& $2.807$	& $2.653$		& $6.897$	\\
\textbf{Lorentzian} 	 						
&	& $\eta_{0}$	& $0.269$	& $0.428$	& $2.320$		& $2.913$	& $2.711$		& $7.049$ \\
\hline\hline
\end{tabular}}
\label{tab:am_leffv}
\end{table}

It is of interest to discuss the real space viscosity kernels for monatomic and diatomic systems from a structural point of view. For this purpose we define a structural normalization factor 
\begin{equation}
\xi_{g}=\frac{\int\limits_{0}^{\infty} r [g(r)-1]^2 dr}{\int\limits_{0}^{\infty}[g(r)-1]^2 dr} \label{eqn:norm_g_r}
\end{equation}
where $g(r)$ is the radial distribution function (RDF). Eq.~(\ref{eqn:norm_g_r}) is a measure of the range over which the correlation function decays to zero and therefore could be regarded as a correlation length of the radial distribution function. The RDF (or structure factor in reciprocal space) can be defined either in terms of the vector norm $r_{ij}$ between the atoms $i$ and $j$ or between the centres of mass of molecules $i$ and $j$: $g(r)=\Big\langle \frac{\sum_{i=1}^{N}\sum_{j>1}^{N} \delta(|\mathbf{r}-\mathbf{r}_{ij}|)}{4\pi r^{2}N\rho} \Big\rangle$, where $N$ is the total number of atoms or molecules, and $\rho$ is the atomic or molecular number density. 

The radial distribution functions and normalization factors are presented in figure~\ref{fig:g_r_ad}. We can see that the RDF are typical monatomic and diatomic Lennard-Jones pair correlation functions. $\xi_{g}$ generally increases as we increase the density and temperature from $0.605$ at state point $\rho_{a}=0.375$, $T=0.756$ to $0.730$ at $\rho_{a}=0.480$, $T=1.0$ and only slightly increases as we increase the cutoff distance, i.e. switch from WCA system to a LJ system at the same state point. $\xi_{g}$ for chlorine at state point $\rho_{a}=1.088$, $T=0.97$ was found to be $0.585$.

The normalized kernels with respect to $\eta(y=0)$ and normalization factor $\xi_{g}$ are shown in figures~\ref{fig:etay_2_2}(a) and \ref{fig:etay_2_2}(b). While generally the width of unnormalized kernels increases as we increase the density (figures~\ref{fig:etay_2_1}(a) and \ref{fig:etay_2_1}(b)), the width of the normalized kernels of WCA fluids decreases marginally as we increase the density from $0.376$ (continuous line) to $0.840$ (short-dashed line). The LJ system shows a slightly narrower kernel (dotted line) compared to the WCA system at the same state point (short-dashed line). Though the kernels obtained from both functional forms are quite close to each other (almost identical for values of $y$ of about half of the atomic or molecular diameters, i.e. $y=0.5\sigma$), we can see in figure~\ref{fig:etay_2_2}(a) and \ref{fig:etay_2_2}(b) that the structural normalization did not completely remove the discrepancy between the normalized kernels of the WCA system at different densities, and the normalized kernels of WCA, LJ and chlorine systems, for values higher than $y=\sigma$. If we recall that figure~\ref{fig:etay_2_2}(b) is based on a Lorentzian-type fit, a further question as to whether the kernel differences are due to numerical analysis, i.e. the choice of the fitting function or due to improper structural factor arises. A four-term Gaussian only shows a slightly narrower kernels which suggests a need for a more comprehensive structural normalization. 

\section{Conclusion\label{sect:V}}
The wave-vector dependent viscosity of monatomic Lennard-Jones, monatomic Weeks-Chandler-Andersen and diatomic (liquid chlorine) fluids over a large wave-vector range and for a variety of state points has been computed. The equilibrium molecular dynamics calculation involved the evaluation of the transverse momentum density and shear stress autocorrelation functions in both atomic and molecular hydrodynamic representations for molecular fluids. The main results can be summarized as follows:

(i) For monatomic fluids the shape of the normalized viscosity kernel in reciprocal space in the low wave-vector region is the same for a whole range of densities considered here. Though the normalized reciprocal kernels insignificantly decreases with the density they show a similar limiting behaviour at high $k_{y}$ values. For the LJ potential compared to a WCA potential we find higher viscosities in the low wave-vector region but the normalized shape of the kernels are almost identical. 

(ii) For liquid chlorine, the wave-vector dependent viscosity shows a similar behaviour in both atomic and molecular formalisms within statistical uncertainty. 

(iii) While a relatively simple Lorentzian-type function fits the atomic and diatomic data well over the entire range of $k_{y}$ at all the state points it is not possible to analytically inverse Fourier transform it to the real space domain. Therefore one may consider an expansion up to a four-term Gaussian which gives better accuracy in reciprocal space compared to the Lorentzian-type function. Our analysis of the high $k_{y}$ regime reveals that the two-term equally weighted Gaussian functional form is inaccurate in predicting the real space kernels whilst the unequally weighted Gaussian only slightly improves the fit.

(iv) The overall conclusion is that the real space viscosity kernel for chlorine has a width of roughly 4-6 atomic diameters while for monatomic systems at high densities the width is about 3-5 atomic diameters and 2-4 atomic diameters at low densities. This means that generalized hydrodynamics must be used in predicting the flow properties of molecular fluids on length scales where the gradient in the strain rate varies significantly on these scales. Consequently a nonlocal constitutive equations should be invoked for a complete description of flows at atomic and molecular scales under such conditions.

Finally, our results for molecular fluids should also provide a good test for more complex molecular systems and the methodology can easily be used for instance in chain-like molecules.


\end{document}